\documentclass[onecolumn,12pt]{article}

\usepackage{jheppub}
\usepackage{ifpdf}
\usepackage[bb=boondox]{mathalfa}

\graphicspath{ {figures/} }

\usepackage[svgnames]{xcolor}
\usepackage{amsfonts}
\usepackage{latexsym}
\usepackage{color}
\usepackage{bm}

\usepackage{amsmath,braket}
\usepackage{amssymb,physics}
\usepackage{amsthm}
\usepackage{mathtools}

\usepackage{epsfig}

\usepackage{tensor}
\usepackage{pdfpages}
\usepackage{bbm}
\usepackage{graphicx,epstopdf}
\usepackage[numbers]{natbib} 
\usepackage[makeroom]{cancel}
\usepackage{hyperref}
\usepackage{array}
\usepackage[export]{adjustbox}

\usepackage[normalem]{ulem}
\usepackage{romannum}

\usepackage{ulem}

\usepackage{tikz}
\usetikzlibrary{intersections}
\usepackage{circuitikz}

\usepackage{subcaption}

\numberwithin{equation}{section}									% equation numbering by section

\usepackage{here}

\def \d {\partial}
\def \lll {\langle\!\langle}
\def \rr {\rangle\!\rangle}

\subheader{\today}
\title{\boldmath Revisiting the $k$-theorem with the ANEC}

\author[a]{Nanami Nakamura,}
\author[a]{Yu Nakayama,}
\author[a]{Ung Nguyen}

\affiliation[a]{Center for Gravitational Physics and Quantum Information, Yukawa Institute for Theoretical Physics, Kyoto University,\\
	Kitashirakawa Oiwakecho, Sakyo-ku, Kyoto 606-8502, Japan}
\emailAdd{nanami.nakamura@yukawa.kyoto-u.ac.jp}
\emailAdd{yu.nakayama@yukawa.kyoto-u.ac.jp}
\emailAdd{nguyen@yukawa.kyoto-u.ac.jp}

\abstract{The fundamental theorem in renormalization group flows in two dimensions is the $c$-theorem, which dictates that the number of degrees of freedom must decrease monotonically along the renormalization group flow. The $k$-theorem claims that the number of charged degrees of freedom also decreases monotonically. Here, $k$ is the current central charge defined by the two-point function of the current. A recent derivation of the $c$-theorem by Hartman and Mathys, which uses the three-point function sum rule and the positivity of the averaged null energy (ANE) operator, motivates us to seek a similar proof of the $k$-theorem. In the case of the $k$-theorem, the partial contact terms need to be taken into consideration. While ignoring the partial contact terms yields contradictory results, our careful analysis incorporating them leads to the correct sum rule and a complete proof based on the positivity of the ANE operator.}

\begin{document} 
	
%%%%%%%%%%%%%%%%%%%%%%%%%%%%%%%%%%%%%%%%%%%%%%%%%%%%
\begin{flushright}
YITP-25-177
\\
\end{flushright}
%%%%%%%%%%%%%%%%%%%%%%%%%%%%%%%%%%%%%%%%%%%%%%%%%%%%
\maketitle
\flushbottom

\section{Introduction}
Counting is a fundamental activity for human beings, and so is counting degrees of freedom for physicists. The very first thing to know about a given field theory is how many degrees of freedom it contains. Historically, various measures have been proposed, such as thermal free energy, trace anomaly, or entanglement entropy. Among these, quantities that decrease monotonically along the renormalization group (RG) flow are of particular interest. The $c$-function \cite{Zamolodchikov:1986gt}, $F$-function \cite{Jafferis:2011zi,Klebanov:2011gs,Casini:2012ei}, and $a$-function \cite{Cardy:1988cwa,Komargodski:2011vj} in $d=2,3,4$ dimensions are such examples. The establishment of these theorems has been a cornerstone in theoretical physics.

Various approaches to proving this monotonicity (specifically for the $c$-theorem) exist. In this paper, we will focus on quantum field theories in two dimensions. Even in two dimensions, there are various proofs: the study of two-point functions of the energy-momentum tensor by Zamolodchikov using the positivity of the two-point functions \cite{Zamolodchikov:1986gt}, the study of the entanglement entropy on a segment by Casini and Huerta using the strong subadditivity \cite{Casini:2006es}, or the study of the three-point functions of the energy-momentum tensor using the positivity of the averaged null energy (ANE) operator by Hartman-Mathys \cite{HM2d}. Each of these approaches provides a different viewpoint on what we mean by “counting the degrees of freedom” in field theories.

A more refined question is to count degrees of freedom charged under certain symmetries. Recently, there has been considerable attention to the symmetry-resolved entanglement entropy \cite{ Goldstein:2017bua}, which captures how entanglement is distributed among different charged sectors (see e.g. \cite{Castro-Alvaredo:2024azg} for a review). While the total entanglement entropy decreases along the RG flow (in the case of the sphere entangling surfaces), the symmetry-resolved entanglement entropy may not monotonically decrease along the RG. The value of the symmetry-resolved entanglement entropy at the fixed point, however, does decrease if we compare the ones at the infrared (IR) fixed point and the ultraviolet (UV) fixed point if we believe the entanglement equipartition \cite{2018PhRvB..98d1106X}. This monotonicity, however, is less illuminating because it does not give us more interesting bounds on what can happen along the RG flow as long as the symmetry is preserved. 

In two dimensions with continuous $U(1)$ symmetry, the so-called $k$-theorem gives more non-trivial information on how much the charged degrees of freedom can monotonically decrease in a certain measure along the RG flow \cite{Ludwig94}. So far, the only known proof is based on the study of two-point functions of the current operator. In this paper, we give a different proof based on the study of the three-point functions of the energy-momentum tensor and the current from the positivity of the averaged null energy (ANE) operator,\footnote{The positivity of the ANE operator was proved in \cite{ANECFaulkner:2016mzt,ANECHartman:2016lgu}. More general light-ray operators and their physical applications are discussed, for example, in \cite{Kravchuk:2018htv,Kologlu:2019mfz}.} which is similar to the proof of the $c$-theorem by Hartman-Mathys \cite{HM2d}.

Our naive expectation was that we could repeat their analysis verbatim to derive the $k$-theorem. Surprisingly, this had given us the wrong sign. It turns out that the partial contact terms,\footnote{Partial contact terms are also called semi-local terms \cite{Nakayama:2019mpz}. Their interesting applications can be found in reference \cite{pcBzowski:2014qja,pcDymarsky:2014zja,pcGomis:2015yaa,pcGomis:2016sab,pcBzowski:2017poo,pcSchwimmer:2018hdl,pcCordova:2019jnf} } which did not contribute to the three-point sum rule in the case of the $c$-theorem, do contribute to the sum rule. Crucially, we find their contribution is exactly minus twice that of the non-partial contact term. Eventually, this saved the sign error and enabled us to complete the proof; however, it makes the relation between monotonicity and positivity less intuitive.

It would be interesting if we could give a third proof based on the (symmetry-resolved) entanglement entropy. It would shed some more light on the nature of positivity. This will be left as future work.

The plan of the paper is as follows.
In section 2, we will first review the content of the theorem and its original proof. In section 3, we will give a new derivation for a sum rule involving the three-point function $TJJ$ with an emphasis on the importance of the partial contact terms. In section 4, we will discuss some technical aspects and possible generalizations of the setup.

\subsection*{Notation}
Our Euclidean field theory notation is based on the textbook by Polchinski \cite{Polchinski:1998rq} with $\alpha'=2$.
In particular, we use
\begin{equation}
        z = i\tau + y, \
        \bar{z} = -i\tau +y, 
\end{equation}
with $y$ being the spatial coordinate and $\tau$ being the Euclidean time.
We also use the derivatives: $\d \equiv \frac{\partial}{\partial z} = \frac{1}{2}(\d_y - i \d_\tau)$ and $\bar{\d} = \frac{\partial}{\partial \bar{z}} = \frac{1}{2}(\d_y + i\d_\tau)$.  With the corresponding momentum, the exponent in the Fourier transform becomes $iz p_z + i\bar{z} p_{\bar{z}} = iy p_y + i\tau p_\tau$. The integration measure  becomes $dz d\bar{z} = 2 d\tau dy  = 2d^2 x$ and $dp_z dp_{\bar{z}}= \frac{1}{2} dp_y dp_\tau$. 
Our convention of the energy-momentum tensor again follows the one used in Polchinski's book:
\begin{equation}
    T_{\mu \nu} = -\frac{4\pi}{\sqrt{g}} \frac{\delta S}{\delta g^{\mu \nu}}.
\end{equation}
Note that not only the numerical factor of $2\pi$ but also the overall sign is different from the one used in Hartman-Mathys. 

To discuss the Lorentzian correlation functions, we will adopt the following prescription for analytical continuation
\begin{equation}
    -z \to  u = t-y, \
    \bar{z} \to  v = t+y.    
\end{equation}
The vectors $J, \bar{J}$ are understood as $J_z, J_{\bar{z}}$ respectively.
Under analytic continuation, $J, \bar{J}, \d,\bar{\d}$ are related to $J_u, J_v, \d_u, \d_v$ with a minus sign for the $u$ component i.e. $J \to -J_u, \bar{J} \to J_v, \d \to -\d_u, \bar{\d} \to \d_v$. 
Note that while $J \to -J_u$, we have $JJ \to J_u J_u$ because the minus sign appears twice. 
Likewise, $T = T_{zz}$ and is analytically continued to $T_{uu}$.

Furthermore, we will also adopt the following definitions for momentum-space correlation functions.
\begin{equation}
\begin{split}
    \expval{O_1(q_1)\dots O_n(q_n)} &\equiv (2\pi)^2 \delta^{(2)}(q_1+\dots+q_n)\lll O_1(q_1) \dots O_n(q_n) \rr  \\
    &\equiv \int d^2x_1 \dots d^2 x_n \ e^{i(q_1x_1+\dots + q_nx_n)} \expval{O_1(x_1)\dots O_n(x_n)}.
\end{split}
\end{equation}

Whenever the distinction is necessary, we use the capital letters, such as $Q_\mu$, to denote the Euclidean momentum, while we use the lowercase letters, such as $q_\mu$, to denote the Lorentzian momentum.

In the Lorentzian signature, we will be considering both the time-ordered and retarded three-point functions.
Our definition of the retarded correlation function is taken from \cite{HM4d} or \cite{Meltzer:2021bmb}.
Away from where the insertion points collide, the retarded three-point function is given by
\begin{equation}
\begin{split}
    \expval{\mathcal{R}[ O(x); O(x_1)  O(x_2)]}  = -\bigg(\theta&(x^0-x^0_1) \theta(x^0_1-x^0_2) \expval{[[O(x), O(x_1)], O(x_2)]} \\ &+ \theta(x^0-x^0_2) \theta(x^0_2-x^0_1) \expval{[[O(x), O(x_2)], O(x_1)]}\bigg). 
\end{split}
\end{equation}
We will be interested in analyzing the contact terms, that is, the contribution to the correlation function when a subset of insertion points collide.
The correlation function diverges as the insertion points collide, so such contributions should be thought of as distributions.
In the case we are considering, such terms will be derivatives of Dirac delta functions.

In momentum space, such terms are polynomials in momenta, so they can be analytically continued in an easy manner
\begin{equation}
    \begin{split}
    \lll O(Q) O(Q_1) O(Q_2)\rr_{\text{contact}} &\to -\lll\mathcal{T} [O(q) O(q_1) O(q_2)]\rr_{\text{contact}} \\ &= \lll \mathcal{R} [O(q); O(q_1) O(q_2)]\rr_{\text{contact}}.
    \end{split} \label{continuecontact}
\end{equation} 
Assuming there are no singularities, the position-space Euclidean correlation function is analytically continued to the time-ordered Lorentzian one through $\tau \to it$.
When moving to momentum space, there is an extra factor of $i$ coming from each one of the Fourier transform integrals, so under analytic continuation, there will be an extra factor of $i^{n}$ in $n+1$ point functions,  which explains the first sign in \eqref{continuecontact}.

On the other hand, the relation between retarded and time-ordered correlation functions (for the contact terms) can be understood as follows: from the condition that the spectrum only contains physical momenta, one can argue that 
\begin{equation}
    \expval{\mathcal{T} \ [O(q) O(q_1) \dots O(q_n)]} = i^n\expval{\mathcal{R} \ [ O(q); O(q_1) \dots O(q_n)]}
\end{equation}
whenever all sums of different momenta are spacelike.
The prefactor here comes from how the retarded correlation function was defined as in \cite{Meltzer:2021bmb}. We want to claim that this equality is true in any momentum configuration. The difference starts when the momentum-space correlation functions show a singularity (e.g., a cut or pole). For contact terms, it is an entire function by assumption, which explains the second sign and equality in \eqref{continuecontact}. Note that these simple relations between time-ordered and retarded correlation functions may not be true in more general correlation functions, such as partial contact terms, and a more elaborate analysis will be needed. 
For a more general and formal discussion of the definitions and properties of such correlation functions, the reader is referred to \cite{Meltzer:2021bmb}.

\section{The review of $k$-theorem and generalization}
In this section, we will review the $k$-theorem as was proved in \cite{Ludwig94} and check its validity in simple examples.
The statement of the $k$-theorem is very similar to the $c$-theorem (see e.g. \cite{klecture}).
In the $c$-theorem, we define a $c$-function based on a two-point function of the energy-momentum tensor.
This function decreases along an RG flow and can be interpreted as the number of degrees of freedom at a certain scale.

Before we proceed, let us comment on the physical significance of $k$. The number $k$ has a definite physical meaning only if we fix the normalization of the current $J$ from some principle.
In the original derivation \cite{Ludwig94}, $k$ was defined by a $U(1)$ current that was embedded within an $SU(2)$ symmetry.
More generally, for a non-Abelian symmetry, we have the OPE
\begin{equation}
    j^a(z) j^b(0) \sim \frac{k^{ab}}{z^2} + \frac{i f^{ab}_c j^c(0)}{z},
\end{equation}
and the normalization for $k$ is fixed from the canonically normalized structure constant $f_c^{ab}$. 

One might then say that for purely Abelian symmetries, the normalization $k$ is not physical.
However, as was argued in \cite{Lin:2019kpn}, $k$ is physical whenever the $U(1)$ symmetry is not an $\mathbb{R}$ symmetry and acts faithfully on local operators. Thus, if we can identify a charged operator that remains along the RG flow, the normalization of $k$ should also be fixed and be physical. 
In the following, we implicitly assume that the normalization of $J$ is fixed by the existence of such a charged operator along the renormalization group flow.\footnote{More technically, this condition does not fix the normalization of the current if the local operator is charged under multiple $U(1)$ symmetries. In most of our discussions, we restrict ourselves to the case when we have only a single $U(1)$ symmetry. The case with multiple $U(1)$ symmetries can be more delicate. See further discussions in section \ref{summary}.}

Since $c$ -- appearing in the $TT$ OPE -- is thought of as the number of degrees of freedom, $k$ can be thought of as the number of charged degrees of freedom. Note that in the case of $c$, the normalization of the energy-momentum tensor was fixed
by its role as the generator of spacetime symmetries (via the Ward identities for translation, rotation, and dilatation) acting on the local operators so that $c$ defined by the two-point function of such a normalized energy-momentum tensor has a definite physical significance. 

\subsection{Statement and original proof}
The starting assumption for the $k$-theorem is that there is a $U(1)$ symmetry.
Correspondingly, we have a conserved current which has components $J, \bar{J}$.
For now, we will assume that the (vectorial)\footnote{The name ``vectorial" is just a convention. It can be a chiral current. This adjective is put to make a distinction with the ``axial" current with the flipped sign in the right-mover.} current is conserved along the RG flow
\begin{equation}
    \partial \bar{J} + \bar{\partial} J = 0, \label{conservation}
\end{equation}
and from that equation, one also has the equation
\begin{equation}
    \partial (\partial \bar{J}) + \bar{\partial} (\partial J) = 0. \label{dj:conserved:eq}
\end{equation}
This equation plays the same role as the conservation law in Zamolodchikov's proof of the $c$-theorem; that is, we obtain energy conservation if we replace $\d J$ with $T$ and $\d \bar{J}$ with $\Theta$, the trace of the energy-momentum tensor. 

Note that at the conformal fixed points, if the theory is unitary and has a discrete spectrum, \eqref{conservation} implies that both the left-moving current and the right-moving current are conserved\footnote{When the theory is not unitary or not compact, this may not be the case. See an example in section 4.1.}
\begin{equation}
        \d \bar{J} = 0, \ \ 
        \bar{\d} J = 0,
\end{equation}
and hence the components are holomorphic and antiholomorphic. 
This means that at the fixed point, the axial current, which could be identical to the vectorial current, is also conserved:  $\partial (\partial \bar{J}) - \bar{\partial} (\partial J) = 0$.
We stress, however, that along the RG flow, we do not assume that the axial current is conserved.

Let us prove the $k$-theorem. We introduce our central quantity, the current central charge $k$.
In the UV limit, from conformal symmetry, we have
\begin{equation}
    \expval{J(z)J(0)} = \frac{k}{z^2}
\end{equation}
where $k$ is a positive number.\footnote{The reflection positivity demands that for a Hermitian operator $O(z)$ with integer spin $j$, $\langle O(z) O(0) \rangle = \frac{(-1)^j c_O}{z^{\Delta + j} \bar{z}^{\Delta-j}}$ with positive $c_O$. Our normalization here corresponds to an anti-Hermitian $J$. \label{footnote:ref}} 
It is called the current central charge because it appears as a central term in the Kac-Moody algebra. 
In what follows, we will demonstrate that if the theory flows to a CFT in the IR, then $k$ will always decrease and become zero in the case where the IR theory is gapped.
Following the idea in deriving the $c$-theorem, we define the following two-point functions 
\begin{align} \label{FGH_conserved_eq1}
         F(z\bar{z}\Lambda^2) &= z^4\expval{\d J(z, \bar{z}) \d J(0)},  \\
        G(z\bar{z}\Lambda^2) &= z^3\bar{z}\expval{\d \bar{J}(z,\bar{z}) \d J(0)}, \label{FGH_conserved_eq2}
         \\
        H(z\bar{z}\Lambda^2) &= z^2\bar{z}^2\expval{\d \bar{J}(z,\bar{z}) \d \bar{J}(0)} . \label{FGH_conserved_eq3}
\end{align}
Here $\Lambda$ is the UV cutoff scale. 
By definition, the value of the function $F$ at the UV conformal fixed point and the one at the IR conformal fixed point will become $-6k_{\mathrm{UV}}$ and $-6k_{\mathrm{IR}}$ respectively.\footnote{Here, we assume that the UV and IR fixed points are conformal \cite{Polchinski:1987dy,Nakayama:2013is}.} 
For any scalar function of the form $f(z\bar{z} \Lambda^2)$, we have the relations $\dot{f} \equiv \Lambda^2\d_{\Lambda^2} f = z \d f = \bar{z} \bar{\d} f$.
Using this, we can evaluate the scale derivative on $F, G, H$ as
\begin{equation}
    \begin{split}
        \dot{F} &= z^4\bar{z} \expval{\bar{\d}\d J(z,\bar{z}) \d J(0)} \\
        \dot{G} &= z^4 \bar{z} \expval{\d \d \bar{J}(z,\bar{z}) \d J(0)} + 3z^3 \bar{z} \expval{\d \bar{J}(z,\bar{z}) \d J(0)} \\
        &= z^3 \bar{z}^2 \expval{\bar{\d}\d J(z,\bar{z}) \d \bar{J}(0)} + z^3 \bar{z} \expval{\d J(z,\bar{z}) \d \bar{J}(0)} \\
        \dot{H} &=  z^3\bar{z}^2 \expval{\d \d \bar{J}(z,\bar{z}) \d \bar{J}(0)} + 2z^2\bar{z}^2 \expval{\d \bar{J}(z,\bar{z}) \d \bar{J}(0)}.
    \end{split}
\end{equation}
To get the second line in the expression for $\dot{G}$, we have used the fact that $\expval{\d \bar{J}(z, \bar{z}) \d J(0)} = \expval{\d J(z,\bar{z}) \d \bar{J}(0)}$, which follows from translational and rotational symmetry.	
If we use the conservation equation, $ \partial (\partial \bar{J}) + \bar{\partial} (\partial J) = 0$, we obtain the relations
\begin{align}
    \dot{F} +\dot{G} &= 3G \label{Frel:eq1}\\
    \dot{G} + \dot{H} &= G + 2H. \label{Frel:eq2}
\end{align}
We define the $K$-function: 
\begin{align}
    K \equiv -\frac{1}{6} (F-2G - 3H).
\end{align}
From \eqref{Frel:eq1} and \eqref{Frel:eq2}, we can conclude $\dot{K} = H$. Since $J$ is anti-Hermitian (see footnote \ref{footnote:ref}), the reflection positivity tells $H\leq 0$ and we see that the function $K$ is monotonically decreasing 
\begin{align}
    \dot{K} = H \leq 0.
\end{align}

If we assume $J$ has a canonical scaling dimension as a conserved current at the fixed point, we have $H=0$ because otherwise the two-point functions become scale dependent. The state-operator correspondence then tells $\partial \bar{J}=0$ in compact and reflection positive CFTs. This then implies $G=0$ and the fixed point value of $K$ is nothing but the current central charge $k$ of the fixed point CFT.

As a detour, let us mention the 't Hooft anomaly matching in this context. When we assume the (vectorial) $U(1)$ symmetry is conserved along the RG flow, the 't Hooft anomaly of the (vectorial) $U(1)$ symmetry is given by $k-\bar{k}$. Here $\bar{k}$ is defined by
\begin{align}
\langle \bar{J}(\bar{z}) \bar{J}(0) \rangle = \frac{\bar{k}}{\bar{z}^2}.     \end{align}
Anomaly matching dictates $k-\bar{k}$ must be the same in the UV and IR CFTs.

Within our context, this claim can be easily proved (even without assuming reflection positivity as expected from the nature of the anomaly matching).  Let us define the functions 
\begin{align} \label{FGHbar_conserved}
         \bar{F}(z\bar{z}\Lambda^2) &= \bar{z}^4\expval{\bar{\d} \bar{J}(z, \bar{z}) \bar{\d} \bar{J}(0)},  \\
        \bar{G}(z\bar{z}\Lambda^2) &= \bar{z}^3 z\expval{\bar{\d} J(z,\bar{z}) \bar{\d} \bar{J}(0)}, 
         \\
        \bar{H}(z\bar{z}\Lambda^2) &= z^2\bar{z}^2\expval{\bar{\d} J(z,\bar{z}) \bar{\d} J(0)} .
\end{align}
Let us define $\bar{K}$: $\bar{K} = -\frac{1}{6}(\bar{F}-2\bar{G}-3\bar{H})$ which becomes the antiholomorphic current central charge $\bar{k} = -\frac{1}{6}\bar{F} $ at the fixed points. Taking the scale derivative as above, we obtain $\dot{\bar{K}} = \bar{H}$.

Here is the crucial observation. Due to the conservation law, we find $\bar{H} = H$. Then it means that $K-\bar{K}$ must be a constant along the RG flow, which completes the proof of the 't Hooft anomaly matching. We should emphasize that we only used conservation and not reflection positivity.

In the case of the energy-momentum tensor, we can define the monotonically decreasing antiholomorphic $\bar{c}$-function and show that $c-\bar{c}$ is constant under renormalization, which is the statement that the gravitational anomaly remains constant throughout the RG flow.
For a derivation of such a result, see, for example, appendix D of \cite{Hellerman:2021fla}.

Here, let us comment on the Sugawara construction \cite{Sugawara:1967rw}. 
When a CFT has a continuous symmetry (or conserved currents), one can construct an energy-momentum tensor out of bilinear combinations of symmetry currents.
For the case of a $U(1)$ symmetry with holomorphic current $J$ with current central charge $k$, the corresponding energy-momentum tensor is given by
\begin{equation}
    T^{U(1)} = \frac{1}{2k} :JJ:,
\end{equation}
where the normalization is so that $J$ has the conformal dimension $1$ when one takes the $T^{U(1)} \times J $ OPE.
One then sees that the $U(1)$ contribution to the central charge is always $c^{U(1)} = 1$.
Assuming in the UV, one has such a $U(1)$ symmetry, then the $U(1)$ part of the central charge in the IR will either be $0$ or $1$ (our examples will be of the former case).

In the case where $c^{U(1)}$ goes from $1$ to $0$, the symmetry is trivial in the IR so $k = 0$. On the other hand, it is still possible that the $k$ in the IR remains non-zero while being consistent with the $c$-theorem. This is because the total energy-momentum tensor is given by the sum of the $U(1)$ Sugawara part and the other part: $T^{\mathrm{total}} = T^{U(1)} + T^{\mathrm{other}}$, and the combination of the $c$-theorem and $k$-theorem suggests that while the $c^{U(1)}$ remains the same, $c^{\mathrm{other}}$ will decrease.

\subsection{Generalization to the case when the current is not conserved}
Let us now discuss a small generalization of the $k$-theorem.
So far, we have assumed that the current $J$ is conserved along the RG flow.
We now relax the assumption that the current is conserved along the RG flow, and we only assume that the current is conserved at the UV fixed point and the IR fixed point so that $k$ remains well-defined. 

Suppose $O_{\mu \nu}$ is an operator such that $\d_\mu J_\nu + O_{\mu \nu} = 0$, where $O_{\mu\nu}$ is anti-Hermitian as was $J_\mu$. 
The previous $F, G, H$ functions are redefined as follows:
\begin{align}
\label{nocon}
        F &= z^4 \expval{\d J (z, \bar{z}) \d J(0)} \\
        G &= z^3 \bar{z} \expval{\d J(z, \bar{z}) O_{\bar{z} z}(0)}  \\
        H &= z^2 \bar{z}^2 \expval{O_{\bar{z} z} (z, \bar{z}) O_{\bar{z} z} (0)}. \label{nocon2}
\end{align}
If we take 
\begin{align}
\label{eq:KX}
    K \equiv -\frac{1}{6}(F -2G -3H),
\end{align}
then, following the calculations outlined above, we can show 
\begin{equation}
    \dot{K} = H.
\end{equation}
Let us recall $O_{\mu\nu}$ is anti-Hermitian, and the reflection positivity tells us $H \leq 0$. It implies $\dot{K} \leq 0$ and we obtain the $k$-theorem.
Note that the obvious choice of $O_{\bar{z}z} = -\bar{\d} J$ works here, so apparently we do not need conservation of current for the $k$-theorem.\footnote{We should realize that this is a subtle statement because for our discussions, it is crucial to assume there is a charged operator under the symmetry to fix the normalization to make the theorem physically meaningful. If the symmetry were completely broken along the RG flow and the RG flow were non-perturbative so that the operator identification becomes difficult, it would not be immediately obvious how we can fix the normalization of the IR current even if it is somehow ``recovered" in the IR limit. In our explicit examples below, one of the vectorial or axial currents is preserved, so this issue will not arise.} 

Finally, let us observe that based on the relation $\dot{K} = H $, we can derive the Zamolodchikov type sum rule:
\begin{equation}
\label{2pt_sum}
    k_{\mathrm{UV}} - k_{\mathrm{IR}} = -\frac{1}{\pi} \int d^2x x^2 \expval{O_{\bar{z}z}(x) O_{\bar{z}z} (0)}_{\mathrm{sep}}
\end{equation}
Note that the right-hand side (RHS) should be understood as the separated two-point function. 
That is, the integral has a UV cutoff (say, the integration range is from some radius $\epsilon$ to infinity) and the cutoff radius $\epsilon$ is taken to zero at the end without picking up the contact term contribution.

\subsection{The free boson as an example}\label{boson2}
It may be instructive to show explicit examples of the $k$-theorem. In this subsection and the next one, we explicitly show the monotonicity in the free real boson theory and the free Dirac fermion theory.
Let us start with a free (massive) boson as an example. The Euclidean action is given by
\begin{equation}
    S = \frac{1}{8\pi} \int d^2x \left( \ \d_\mu X \d^\mu X + m^2 X^2 \right) = \frac{1}{4\pi} \int dzd\bar{z} \ \left(\d X \bar{\d} X + \frac{1}{4}m^2 X^2 \right).
\end{equation}
The massive scalar field propagator is given by the modified Bessel function as
\begin{align}
\label{scal_prop}
  \expval{X(z, \bar{z}) X(0)}= 2 K_0(m|z|).
\end{align}
When the mass is zero, we have a (vectorial) conserved $U(1)$ current:
\begin{align}
    J=i\d X,\ \ \bar{J}= i\bar{\d}X.
\end{align}

\begin{figure}[h]
  \centering
  \includegraphics[width=0.5\columnwidth]{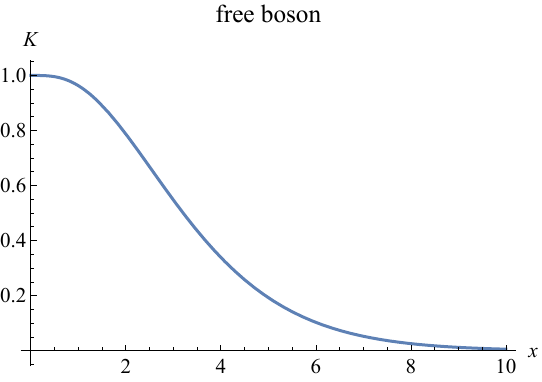}
  \caption{$K$ for the free boson as a function $x=m|z|$}
  \label{fig:K_f0}
\end{figure}

Let us check the $k$-theorem explicitly for this vectorial current $J$. 
Here, because the vectorial current $J$ is not conserved along the RG flow, \eqref{nocon}-\eqref{nocon2} is used with $O_{z\bar{z}}=-\bar{\d} J$.\footnote{Alternatively, we may talk about the axial current $J_A= i\partial X$ and $\bar{J}_A = -i\bar{\partial}X$, which is conserved along the RG flow. We will obtain the same result.} 
We now obtain 
\begin{align}
\begin{split}
     \frac{F(m|z|)}{z^4} &= \ev{\d J(z) \d J(0)} = - 2\d^4 K_0(m|z|) \\
     \frac{G(m|z|)}{z^3\bar{z}} &=-\ev{\d J(z) \bar{\d} J(0)} = 2\d^3\bar{\d} K_0(m|z|) \\
     \frac{H(m|z|)}{z^2\bar{z}^2} &= \ev{\bar{\d} J(z) \bar{\d} J(0)} = -2\d^2\bar{\d}^2 K_0(m|z|),
\end{split} 
\end{align}
and introducing the dimensionless quantity $x = m|z|$, they become
\begin{align}
\begin{split}
    F(x) &= -\frac{1}{8}[x^2(24+x^2)K_0(x) + 8x(6+x^2)K_1(x)]\\
    G(x) &= \frac{1}{8} x^4 K_2(x) \\
    H(x) &= -\frac{1}{8} x^4 K_0(x)
\end{split}
\end{align}
Using \eqref{eq:KX}, we can obtain the following expression for the $K$-function:
\begin{align}
    K(x)=\frac{1}{48} \qty[(-2 x^4+24x^2)K_0(x) + 8x(6+x^2)K_1(x) +2 x^4 K_2(x)].
\end{align}
We plot the function $K$ in Figure \ref{fig:K_f0} to see that $K$ is monotonically decreasing.

\subsection{The free fermion as an example}
\begin{figure}
    \centering
    \includegraphics[width=0.5\linewidth]{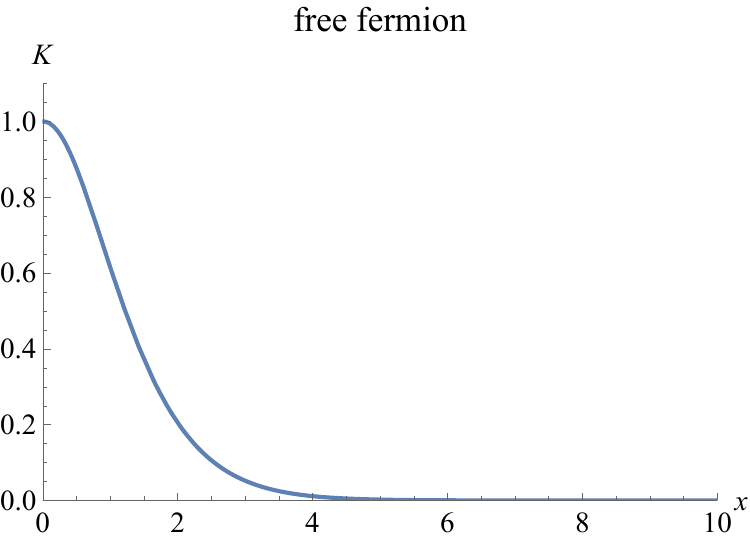}
    \caption{$K$ for the free fermion as function of $x = m|z|$ } 
    \label{fig:fermion2pt}
\end{figure}

We will also show a similar calculation for the free Dirac fermion theory.
Our Euclidean action for the free Dirac fermion is
\begin{equation}
    S = \frac{1}{2\pi}\int dz d\bar{z} \bigg(\psi^\dagger \bar{\partial} \psi + \tilde{\psi}^\dagger \partial \tilde{\psi} - \frac{m}{2}(\tilde{\psi}^\dagger \psi + \psi^\dagger \tilde{\psi}) \bigg).
\end{equation}
We find the two-point function to be
\begin{equation}
\begin{split}
        \begin{pmatrix}
        \expval{\psi(x) \tilde{\psi}^\dagger(y)} & \expval{\psi(x) \psi^\dagger(y)} \\
        \expval{\tilde{\psi}(x) \tilde{\psi}^\dagger(y)} & \expval{\tilde{\psi}(x) \psi^\dagger(y)}
    \end{pmatrix}
    &= - \int \frac{d^2 p}{2\pi} \frac{e^{ip(x-y)}}{p^2+m^2} 
    \begin{pmatrix}
        m & 2ip_z \\
        2ip_{\bar{z}} & m
    \end{pmatrix} \\
    &= 
    -\begin{pmatrix}
     m & 2\d \\
     2\bar{\d} & m
    \end{pmatrix} K_0(m|z|)
\end{split}
\end{equation}
with $K_0$ being the modified Bessel function.
This theory has a conserved vectorial $U(1)$ current (even with the mass term) given by
\begin{equation}
    J = -\psi^\dagger \psi, \ \bar{J} = - \tilde{\psi}^\dagger \tilde{\psi}.
\end{equation}
It is conserved in the sense $\bar{\partial} J + \partial \bar{J} = 0$ along the RG flow while the axial current $    J_A = -\psi^\dagger \psi, \ \bar{J}_A =+ \tilde{\psi}^\dagger \tilde{\psi}$ is broken.
Wick's theorem gives
\begin{equation*}
    \begin{split}
        \frac{F(m|z|)}{z^4}=\expval{\d J(z, \bar{z}) \d J(0)} &= -\d^2 \{ (2\d K_0(m|z|))^2 \} \\
        \frac{G(m|z|)}{z^3\bar{z}}=\expval{\d \bar{J}(z,\bar{z}) \d J(0)} &= \d^2 \{ (m K_0(m|z|))^2 \}\\
        \frac{H(m|z|)}{z^2\bar{z}^2}=\expval{\d \bar{J}(z,\bar{z}) \d \bar{J}(0)} &= -\d^2 \{ (2\bar{\d}K_0(m|z|))^2 \}.
    \end{split}
\end{equation*}
Here, since the current is conserved along the flow, we employed the definitions \eqref{FGH_conserved_eq1}-\eqref{FGH_conserved_eq3}. 
We can then obtain an expression for $K = -\frac{1}{6}(F-2G-3H)$ which reads
\begin{equation}
    \begin{split}
    K(x) &= \frac{x^2}{24} (x^2K_0(x) +18x K_0(x) K_1(x)  \\&+ (8+x^2)K_1(x)^2 - x^2 K_2(x)^2 + xK_1(x)(14K_2(x)-xK_3(x))).
    \end{split}
\end{equation}
Numerically evaluating this, we can see that the function decreases as one goes from the UV to the IR (see Figure \ref{fig:fermion2pt}).

Let us remark here that the (compactified) boson and the Dirac fermion are equivalent under the bosonization duality. The mass term for the Dirac fermion can be mapped to the sine-Gordon potential for the boson. Since we study the conventional mass term $m^2 X^2$ rather than the sine-Gordon potential, the behavior of the $K$ function is slightly different. If we could analyze with the sine-Gordon potential, we would obtain the same plot.

\section{Three-point sum rule and positivity}
In this section, we will show the main results of the paper, that is, a derivation of a sum rule for $k$ and the proof of the $k$-theorem based on the positivity of the ANE operator.
The derivation here follows the same strategy as that of \cite{HM2d} and \cite{HM4d} for the proof of the $c$ and $a$ theorems, but we will find a crucial difference in the contributions from the partial contact terms.

Let us outline the strategy.
First, we identify $k$ as a coefficient of a contact term within the three-point function $\expval{T \bar{\d}J\bar{\d}J }$.
From that, we can relate the difference between $k_{\mathrm{UV}}$ and $k_{\mathrm{IR}}$ as an integral over a separated correlation function.
This is the sum rule.
Based on the sum rule, we then assume that $\Delta k \equiv k_{\mathrm{UV}} - k_{\mathrm{IR}}$ can also be written as the expectation value of the ANE operator
\begin{equation}
    \mathcal{E}(v) \equiv - \int du \ T_{uu}(u, v)
\end{equation}
with respect to a judiciously chosen state.
From the positivity of the ANE operator, which we will also call the averaged null energy condition or ANEC,\footnote{Note that our convention of the energy-momentum tensor has a sign difference from the one used by Hartman-Mathys, so we have put the extra minus sign here to make $\mathcal{E}$ positive.} the $k$-theorem shown above should be reproduced. 

\subsection{An attempt at deriving the sum rule and a puzzle}
In this section, we derive a sum rule of the three-point function and attempt to prove the $k$-theorem  in a way that is similar to \cite{HM2d} and \cite{HM4d}.
To derive the sum rule, our starting point is the contact terms of the correlation function $\expval{\bar{\d}J(x_1) \bar{\d}J(x_2) T(x_3)}$ in a CFT.
In fact, $\bar{\d} J = 0$ in a CFT, so the correlation function should only contain contact terms.
From the current two-point functions $\expval{J(x_1)J(x_2)} = \frac{k}{z_{12}^2}$, we have
\begin{equation}
    \expval{J(x_1)J(x_2) T(x_3)} = \frac{k}{z_{13}^2z_{23}^2},
\end{equation}
which follows from the conformal Ward identity.
Using the identity $\bar{\d} \frac{1}{z} = \pi \delta^{(2)}(x)$, we obtain
\begin{equation}
    \expval{\bar{\d}J(x_1) \bar{\d} J(x_2) T(x_3)} = k \pi^2 \partial_{z_1}\delta^{(2)}(x_{13}) \partial_{z_2}\delta^{(2)}(x_{23}).
\end{equation}
Notice that this expression does not have any partial contact terms.
This derivation using the Ward identity bypasses potentially lengthy calculations involving functional derivatives (See Appendix of Hartman-Mathys \cite{HM2d}).
In momentum space, we have 
\begin{equation}
    \begin{split}
        \expval{T(Q_3) \bar{\d} J(Q_1) \bar{\d} J(Q_2)} &=
        \int d^2 x_1 d^2 x_2 d^2 x_3 e^{i(Q_1x_1 + Q_2 x_2 + Q_3 x_3)} \expval{T(x_3) \bar{\d} J(x_1) \bar{\d} J(x_2)} \\
        &=(2\pi)^2 \delta^{(2)}(Q_1+Q_2+Q_3) \qty(-k \pi^2  Q_{1z} Q_{2z}),
    \end{split}
\end{equation}
or equivalently 
\begin{equation} \label{cft_contact_term:eq}
    \lll T(Q_3) \bar{\d} J(Q_1) \bar{\d} J(Q_2) \rr = -k \pi^2  Q_{1z} Q_{2z}.
\end{equation}
Since this formula for the correlation function is just a contact term, the above Euclidean formula can be easily analytically continued into a Lorentzian signature
\begin{equation}
    \lll \mathcal{R} \qty[ T_{uu}(q_3); \d_v J_u(q_1) \d_v J_u(q_2)] \rr  =  -\lll \mathcal{T} \qty[ T_{uu}(q_3) \d_v J_u(q_1) \d_v J_u(q_2)] \rr 
    % = - \lll \mathcal{T} \ T(q_3) \bar{\d} J(q_1) \bar{\d} J(q_2) \rr 
    = -k \pi^2 q_{1u} q_{2u}.
\end{equation}

We recall that the definition of the ANE operator is given by\footnote{Recall  $d^2q =2 dq_u dq_v$.}
\begin{equation*}
    \mathcal{E}(v = 0) = - \int du T_{uu}(u, v=0) = - \frac{1}{\pi}\int dq_v  T_{uu}(q_u = 0, q_v).
\end{equation*}
Note that the final expression is an integral of the momentum-space representation of the energy-momentum tensor. 
For the remainder of this paper, when the argument of $\mathcal{E}$ is not explicitly shown, it is assumed that $v = 0$.
From the definition and the contact terms that we computed, we obtain\footnote{Recall $\delta^{(2)}(q) = \frac{1}{2} \delta(q_{u})\delta(q_v)$.}
\begin{equation}
\begin{split}
    &\expval{\mathcal{R} \ [\mathcal{E}(v=0);\d_v J_u(q_1) \d_v J_u(q_2)]} \\ = &-\frac{1}{\pi} \int dq_{3v} (2\pi)^2 \delta^{(2)}(q_1+q_2+q_3) \lll \mathcal{R} \ [T_{uu}(q_3); \d_v J_u(q_1) \d_v J_u(q_2)] \rr \\ = & -2 \pi^3 \delta(q_{1u} + q_{2u}) k q_{1u}^2.
\end{split}
\end{equation}
Notice that the introduction of $\mathcal{E}$ is equivalent to setting $q_{3} = 0$ and multiplying $-2\pi$ within the momentum-space correlation function.
To obtain a desired sum rule for the current central charge $k$ in CFT, we simply invert the above expression.
\begin{equation}
\begin{split}
    k &= -\frac{1}{4\pi^3}  \frac{\partial^2}{\partial q_{1u}^2}\bigg|_{q_1  = 0} \int dq_{2u}\expval{\mathcal{R}  [\mathcal{E}(v=0);\d_v J_u(q_1) \d_v J_u(q_2)]}  \\
    &= \frac{1}{2\pi^2} \int d^2 x_1 d^2 x_2 u_1^2 \delta(u_2) \expval{\mathcal{R} [\mathcal{E}; \d_v J_u(x_1) \d_v J_u(x_2)]} \\
    &= \frac{1}{\pi^2} \int d^2 x_1 d^2 x_2 u_1^2 \delta(u_2) \theta(-v_1) \theta(-v_2) \expval{ \d_v J_u(x_1) \mathcal{E} \d_v J_u(x_2)}. 
\end{split}
\end{equation}
Generalizing the above argument to the general QFT, we can identify $k_{\mathrm{IR}}$ with the integral 
\begin{equation}
\begin{split}
    k_{\mathrm{IR}} =& {+} \frac{1}{2\pi^2} \int d^2 x_1 d^2 x_2 \delta(u_2) u_1^2 \expval{\mathcal{R} [\mathcal{E}(v=0); \d_v J_u(x_1) \d_v J_u(x_2)]}_{\mathrm{QFT}} \\
        =& +\frac{1}{2\pi^2} \qty((\d_{q_{1u}} -\d_{q_{2u}})^2 \lll\mathcal{R} [T_{uu}(q_3) ;\d_v J_u(q_1) \d_v J_u(q_2)] \rr_{\mathrm{QFT}} )  \big|_{q_1=q_2=0} .
\end{split}
\end{equation}
We can then separate the three-point function into a contribution from the UV (when the three points coincide), a contribution from the partial contact, or PC term (when two of the three points coincide), and a contribution from the separated correlation functions (when no points coincide)
\begin{equation}
\begin{split}
    \expval{\mathcal{R} [\mathcal{E}(v=0); \d_v J_u(x_1) \d_v J_u(x_2)]}_{\mathrm{QFT}}  &= \expval{\mathcal{R} [\mathcal{E}(v=0); \d_v J_u(x_1) \d_v J_u(x_2)]}_{\mathrm{UV}} \\ &+ \expval{\mathcal{R} [\mathcal{E}(v=0); \d_v J_u(x_1) \d_v J_u(x_2)]}_{\mathrm{PC}} \\ &+ \expval{\mathcal{R}[ \mathcal{E}(v=0); \d_v J_u(x_1) \d_v J_u(x_2)]}_{\mathrm{sep}},
\end{split}
\end{equation} 
or we could have started with the time-ordered correlation functions, in which case we would obtain
\begin{equation}
\begin{split}
    \expval{\mathcal{T} [\mathcal{E}(v=0) \d_v J_u(x_1) \d_v J_u(x_2)]}_{\mathrm{QFT}}  &= \expval{\mathcal{T} [\mathcal{E}(v=0) \d_v J_u(x_1) \d_v J_u(x_2)]}_{\mathrm{UV}} \\ &+ \expval{\mathcal{T} [\mathcal{E}(v=0) \d_v J_u(x_1) \d_v J_u(x_2)]}_{\mathrm{PC}} \\ &+ \expval{\mathcal{T}[ \mathcal{E}(v=0) \d_v J_u(x_1) \d_v J_u(x_2)]}_{\mathrm{sep}}.
\end{split}
\end{equation}
In both cases, the UV contribution can be identified with $k_{\mathrm{UV}}$.
So far, nothing is prone to errors.

To proceed, for the time being, let us (incorrectly -- to be found later) ignore the contribution from the partial contact term, which was justified in the case of the $c$-theorem \cite{HM2d}.
Then, the sum rule for the current central charge $k$ in the QFT would become 
\begin{equation*}
\label{dk_mis}
\begin{split}
    \Delta k &\stackrel{?}{=}  -\frac{1}{2\pi^2} \int d^2 x_1 d^2 x_2 \delta(u_2) u_1^2 \expval{\mathcal{R} [\mathcal{E}(v=0); \d_v J_u(x_1) \d_v J_u(x_2)]}_{\mathrm{sep}} \\
     &\stackrel{?}{=} +\frac{1}{2\pi^2} \int d^2 x_1 d^2 x_2 \delta(u_2) u_1^2 \expval{\mathcal{T} [\mathcal{E}(v=0) \d_v J_u(x_1) \d_v J_u(x_2)]}_{\mathrm{sep}}.
    \end{split}
\end{equation*}
Or, in the momentum space, we would obtain
\begin{align*}
    \Delta k 
        \stackrel{?}=& -\frac{1}{2\pi^2} \qty((\d_{q_{1u}} -\d_{q_{2u}})^2 \lll\mathcal{R} [T_{uu}(q_3) ;\d_v J_u(q_1) \d_v J_u(q_2) ]\rr_{\mathrm{sep}} )  \big|_{q_1=q_2=0} \\
       \stackrel{?}=  &+\frac{1}{2\pi^2} \qty((\d_{q_{1u}} -\d_{q_{2u}})^2 \lll\mathcal{T} [T_{uu}(q_3) \d_v J_u(q_1) \d_v J_u(q_2)] \rr_{\mathrm{sep}} ) \big|_{q_1=q_2=0} .
\end{align*}
We will see that these would-be sum rules have the wrong sign.

To derive the $k$-theorem from the ANEC, we will consider the state
\begin{equation}
    \ket{\Psi(q_u)} = \int d^2x \theta(-v) e^{iq_u u - \frac{u^2}{\sigma^2}}  \d_v J_u(u, v) \ket{0}
\end{equation}
 with $\sigma^{-1} \ll q_u \ll M$, where $\sigma$ is some IR regulator and $M$ is some UV regulator.
 The condition that $q_u \ll M$ is so that we can do a low-momentum expansion later.
Noting that $J$ is anti-Hermitian (so being careful about the sign flip under the conjugation), we would find 
\begin{equation}
    \begin{split}
    &\bra{\Psi(q_u)} \mathcal{E}(v = 0) \ket{\Psi(q_u)}  \\=& - \int_{v_1 < 0, v_2 < 0} d^2x_1 d^2x_2 \exp(-iq_u u_1 + iq_u u_2 - \frac{u_1^2}{\sigma^2} - \frac{u_2^2}{\sigma^2}) \ \expval{\d_v J_u(x_1) \mathcal{E} \d_v J_u(x_2)}
    \end{split}
\end{equation}
This would-be sum rule is equivalent to the statement that at low momentum, we have the expansion
\begin{align*}
    &\int d^2x_1 d^2x_2 e^{iq_1 x_1+ iq_2 x_2} \expval{\mathcal{R} \ [\mathcal{E}; \d_v J_u(x_1) \d_v J_u(x_2)]} 
    \stackrel{?}{\approx}  + 2\pi^3\Delta k \  q_{1u}^2 \delta(q_{1u}+q_{2u}) + O\qty(\frac{q_{1u}^3}{M})
\end{align*}
To leading order in the small momentum limit, we would obtain 
\begin{equation*}
    -\Delta k \sigma C q_u^2 \stackrel{?}{\approx} \bra{\Psi(q_u)} \mathcal{E} \ket{\Psi(q_u)} 
\end{equation*}
with  $C = \frac{\pi^{5/2}}{\sqrt{2}}$ being a positive numerical value.
This, however, would give us a confusing result that $\Delta k = k_{\mathrm{UV}} - k_{\mathrm{IR}} \leq 0$ from the positivity of the ANE operator, which would be in contradiction with the results explained in the previous section.

\subsection{The correct sum rule and partial contact terms}
In the previous section, we encountered an apparent contradiction in the expression for $\Delta k$. This originates from neglecting the partial contact terms. In this subsection, we properly account for the partial contact terms and demonstrate how their inclusion resolves the contradiction.

Inside the $TJJ$ correlation functions, there are two types of partial contact terms. When the two $\bar{\d}J$ coincide, two currents must be ordered on the same side of ANE operator. Since $\mathcal{E}\ket{0}=0$, the $JJ$ partial contact terms do not contribute, so in the following we will not discuss them. On the other hand, it is necessary to examine the partial contact terms where $T$ coincides with one of the currents.

To see its importance, let us study the OPE between $T$ and $J$: 
\begin{equation}
    T(x_3) J(x_2) = \frac{J(x_3)}{z_{32}^2} + O(1).
\end{equation}
Taking the derivative, we obtain $T(x_3) \bar{\d} J(x_2) = -\pi J(x_3) \d_2 \delta^{(2)}(x_{32})$, and it is important to note that there are no other divergences in the OPE as the two points collide. 
This gives a partial contact term that is proportional to $\expval{J(x_3) \bar{\d}J(x_1)}$ that will eventually contribute to the sum rule. This was the main error in the previous subsection.

We shall now evaluate the partial contact terms.
Within an Euclidean CFT, the two-point function of the current gives 
\begin{equation}
   \expval{\bar{\d}J(x_1) J(x_3)} = -\pi k \partial_1 \delta^{(2)}(x_{13}),
\end{equation}
where $k$ is the current central charge in the CFT.
In momentum space, this becomes 
\begin{equation}
    \expval{\bar{\d} J(Q_1) J(Q_3)} = (2\pi)^2 \delta^{(2)}(Q_1+Q_3) i\pi Q_{1z} k.
\end{equation}
Now, within a QFT, we have
\begin{equation}
    \begin{split}
    &\int d^2x_1 d^2x_3 e^{iQ_1x_1+iQ_3x_3} \expval{\bar{\d} J(x_1) J(x_3)} \\ 
    =& \int d^2x_1 d^2x_3 e^{iQ_1x_1+iQ_3x_3} 
    \big(\expval{\bar{\d} J(x_1) J(x_3)}_{\mathrm{sep}} +\expval{\bar{\d} J(x_1) J(x_3)}_{\text{UV}} \big).
    \end{split}
\end{equation}
In the low momentum limit, the left-hand side gives a contribution proportional to $k_{\mathrm{IR}}$.
As a result, we can conclude that
\begin{equation}
    \expval{\bar{\d}J(Q_1) J(Q_3)}_{\mathrm{sep}}=-(2\pi)^2 \delta^{(2)}(Q_1+Q_3) i\pi Q_{1z} \Delta k + O(Q^2).
\end{equation}

Combined with the contribution from $\d \delta^{(2)}(x_{32})$, the partial contact terms as $x_3$ approaches $x_2$ are given by
\begin{equation}
\label{PC}
    \begin{split}
        & -\int d^2x_1d^2x_2 d^2x_3 e^{iQ_1x_1+iQ_2x_2 + iQ_3x_3} \pi\d_2 \delta^{(2)}(x_{23}) \expval{\bar{\d}J(x_1) J(x_3)}_{\mathrm{sep}} \\
        % =& \int d^2x_1d^2x_2 e^{iQ_1x_1 + i(Q_2+Q_3)x_3} (+i\pi Q_{2z}) \expval{\bar{\d}J(x_1) J(x_2)}_{\mathrm{sep}} \\
        =& (2\pi)^2 \delta^{(2)}(Q_1+Q_2+Q_3) (+i\pi Q_{2z}) \lll \bar{\d}J(Q_1) J(Q_2+Q_3)\rr \\
        =& (2\pi)^2 \delta^{(2)}(Q_1+Q_2+Q_3)\Delta k \pi^2 Q_{1z}Q_{2z} + O(Q^3)
    \end{split}
\end{equation}
in the low momentum limit. 
The partial contact term as $x_1$ approaches $x_3$ gives an almost identical contribution.
Overall, we obtain $\lll T(Q_3) \bar{\d}J(Q_1) \bar{\d}J(Q_2) \rr_{\mathrm{PC},TJ} = 2\times \pi^2\Delta k Q_{1z}Q_{2z}$ where the subscript $TJ$ here emphasizes that we only focus on the partial contact terms that appear when $T$ approaches either of the $\bar{\d} J$'s.
Analytically continuing the above expression to the Lorentzian signature, we obtain the partial contact term sum rule:
\begin{equation}
\label{PC_L}
    \begin{split}
  2 \times \Delta k &=  +\frac{1}{2\pi^2} (\d_{q_{1u}} -\d_{q_{2u}})^2 \lll\mathcal{T} \ [T_{uu}(q_3) \d_v J_u(q_1) \d_v J_u(q_2)] \rr_{\mathrm{PC},TJ}|_{q_1 = q_2 =0} .
    \end{split}
\end{equation}

Taking into account these partial contact contributions resolves the aforementioned puzzle as follows. If we had ignored the contribution from the partial contact terms as we did in the previous subsection, we would have arrived at the wrong sum rule:
\begin{equation*}
    \begin{split}
    \expval{\mathcal{T}[\mathcal{E} \d_v J_u(q_1) \d_v J_u(q_2)] }_{\mathrm{sep}}  &\stackrel{?}{=} \expval{\mathcal{T}[\mathcal{E} \d_v J_u(q_1) \d_v J_u(q_2)] }_{\mathrm{QFT}} - \expval{\mathcal{T}[\mathcal{E} \d_v J_u(q_1) \d_v J_u(q_2)] }_{\mathrm{UV}} \\
    &\approx +2\pi^3 \delta(q_{1u} + q_{2u}) \Delta k \ q_{1u} q_{2u}.
    \end{split}
\end{equation*}
in the low momentum limit. 
Recall that in the low momentum limit, the QFT correlation function is identified with that of the IR CFT.
Since the partial contact term also contributes to the sum rule, the correct sum rule should be given by
\begin{equation}
    \begin{split}
    &\expval{\mathcal{T}[\mathcal{E} \d_v J_u(q_1) \d_v J_u(q_2) ]}_{\mathrm{sep}} \\  = & \expval{\mathcal{T}[\mathcal{E} \d_v J_u(q_1) \d_v J_u(q_2)] }_{\mathrm{QFT}} - \expval{\mathcal{T}[\mathcal{E}\d_v J_u(q_1) \d_v J_u(q_2)] }_{\mathrm{UV}} -\expval{\mathcal{T}[\mathcal{E} \d_v J_u(q_1) \d_v J_u(q_2)]}_{\mathrm{PC}} \\
     \approx& -2\pi^3 \delta(q_{1u} + q_{2u}) \Delta k \ q_{1u} q_{2u}.
    \end{split} \label{dec}
\end{equation}

Indeed, we know that the first two terms gave the ``wrong contribution" $ +2\pi^3 \delta(q_{1u} + q_{2u}) \Delta k \ q_{1u} q_{2u}$ because they are nothing but the would-be sum rule which neglects the partial contact terms, where the sign was wrong.
On the other hand, we have just calculated the contributions from the partial contact terms in \eqref{PC_L}, which gives $ -4\pi^3 \delta(q_{1u} + q_{2u}) \Delta k \ q_{1u} q_{2u}$. Thus, combining all together, we conclude that \eqref{dec} gives the correct sum rule in momentum space
\begin{align}
    \Delta k 
    =& +\frac{1}{2\pi^2} \qty(\d_{Q_{1z}} -\d_{Q_{2z}})^2 \lll T(Q_3) \bar{\partial} J(Q_1) \bar{\partial} J(Q_2) \rr_{\mathrm{sep}}) \big|_{Q_1=Q_2=0} \label{dk_Euk} \\
    =&-\frac{1}{2\pi^2} \qty((\d_{q_{1u}} -\d_{q_{2u}})^2 \lll\mathcal{T} [T_{uu}(q_3) \d_v J_u(q_1) \d_v J_u(q_2)] \rr_{\mathrm{sep}} ) \big|_{q_1=q_2=0} \\
        =& +\frac{1}{2\pi^2} \qty((\d_{q_{1u}} -\d_{q_{2u}})^2 \lll\mathcal{R} [T_{uu}(q_3) ;\d_v J_u(q_1) \d_v J_u(q_2)] \rr_{\mathrm{sep}} )  \big|_{q_1=q_2=0} . \label{dk_mom}
\end{align}
Compared with the wrong sum rule in the previous subsection, the entire sign is flipped thanks to the partial contact terms. The Euclidean sum rule \eqref{dk_Euk} is most conveniently used in the examples in the following subsections, while the retarded one is most conveniently used to show its positivity.

To explain the last equality in \eqref{dk_mom}, it is convenient to write down the corresponding position-space sum rule:
\begin{align} 
\label{dk_corr}
    \Delta k &= 
     -\frac{1}{2\pi^2} \int d^2x_1 d^2x_2  \delta(u_2)  u_1^2 \expval{\mathcal{T} [\mathcal{E}(v=0) \d_v J_u(u_1, v_1) \d_v J_u(u_2, v_2)]}_{\mathrm{sep}}  \\
    &=\frac{1}{2\pi^2} \int d^2x_1 d^2x_2 \delta(u_2) u_1^2 \expval{\mathcal{R} [\mathcal{E}(v=0); \d_v J_u(u_1, v_1) \d_v J_u(u_2, v_2)]}_{\mathrm{sep}} \label{dk_corr_R}
\end{align}
They are explicitly related by the rotation of the contour of $v$.\footnote{Since this is a non-trivial point, we show the detailed derivation. While the time-ordered correlation function in \eqref{dk_corr} gives the integral 
\begin{equation}
\label{dk_T}
    \text{\eqref{dk_corr}} = -\frac{1}{\pi^2}\int_{v_1 > 0} d^2x_1 \int_{v_2 < 0} d^2x_2  u_1^2 \delta(u_2) \expval{\d_v J(x_1) \mathcal{E} \d_v J(x_2)}_{\mathrm{sep}},
\end{equation}
the retarded one gives 
\begin{equation}
\label{dk_R}
    \text{\eqref{dk_corr_R}} = \frac{1}{\pi^2}\int_{v_1 < 0} d^2x_1 \int_{v_2 < 0} d^2x_2  u_1^2 \delta(u_2) \expval{\d_v J(x_1) \mathcal{E} \d_v J(x_2)}_{\mathrm{sep}};
\end{equation}
where we have replaced the ordered correlation functions with Wightman functions.
This means that they have different $v_1$ integration contours.
We, however, see that
\begin{equation*}
    \text{\eqref{dk_T}}-\text{\eqref{dk_R}}
    =\int_{-\infty}^\infty dv_1 \int du_1 d^2x_1 \int_{v_2 < 0} d^2x_2  u_1^2 \delta(u_2) \expval{\d_v J(x_1) \mathcal{E} \d_v J(x_2)}_{\mathrm{sep}} = 0
\end{equation*}
because we are integrating over a total derivative of $v_1$ and the Wightman functions do not contain contact terms.
Hence, we can conclude that the two correlation functions only differ by a sign.\label{TR_QFT}} The position-space formulas are just Fourier transforms of the expressions in \eqref{dk_mom}.

Once we have the expression in terms of the retarded correlation functions, we can repeat the positivity argument at the end of the previous subsection; using the wavepacket state defined in the last section, we get
\begin{equation}
    \Delta k \sigma C q_u^2 \approx \bra{\Psi(q_u)} \mathcal{E} \ket{\Psi(q_u)},
\end{equation}
which is almost the same as the previous section, except for a flip in the sign of the LHS.
We conclude that the positivity of $\Delta k$ follows from the positivity of the ANE operator.

\subsection{On the relation between the two-point and three-point sum rule}
Let us revisit the two-point function sum rule.
In \cite{HM2d}, it was demonstrated that the three-point sum rule based on the ANE operator is equivalent to the Zamolodchikov sum rule (the sum rule with the two-point function).
Here, we will show that the two-point function sum rule gives a three-point sum rule with the correct sign. This will reconfirm the validity of the discussions in the previous subsection.

We begin with the two-point sum rule \eqref{2pt_sum}:
\begin{equation}
    \Delta k = -\frac{1}{\pi} \int d^2x  \ x^2 \expval{\bar{\d} J(x) \bar{\d} J(0)}_{\mathrm{sep}}.
\end{equation}
We first Wick rotate the RHS to obtain
\begin{align}
    \Delta k &= -\frac{i}{\pi} \int d^2x  \ x^2 \expval{\mathcal{T} \d_v J_u(x) \d_v J_u(0)}_{{\mathrm{sep}}} 
    = -\frac{2i}{\pi} \int d^2x \ \theta(v) x^2 \expval{\d_v J_u(x)\d_v J_u(0)}_{\mathrm{sep}} .
\end{align}
From this point onwards, all expressions will be in Lorentzian signature.
Here, a step function was introduced to rewrite the time-ordered correlation function into a Wightman function.
Next, using the identity $\int_0^\infty dv  \ v f(v) = \int_0^\infty dv_1 \int_{-\infty}^0 dv_2 f(v_1-v_2)$, we find that the above expression is equivalent to the following:
\begin{align*}
    \Delta k &= \frac{i}{\pi} \int_{-\infty}^\infty du \int dv_1 dv_2 \theta(v_1)\theta(-v_2) u \expval{\d_v J_u(u, v_1) \d_v J_u(0, v_2)}_{\mathrm{sep}} \\
    &= -\frac{i}{2\pi} \int_{-\infty}^\infty du \int dv_1 dv_2 \theta(v_1)\theta(-v_2) u^2 \d_u \expval{\d_v J_u(u, v_1) \d_v J_u(0, v_2)}_{\mathrm{sep}} \\
    &= -\frac{1}{\pi^2}\int d^2x_1 d^2x_2 \theta(v_1) \theta(-v_2) u_1^2 \delta(u_2) \expval{\d_v J_u(x_1) \mathcal{E} \d_v J_u(x_2)}_{\mathrm{sep}}
\end{align*}
Note that here we have replaced the integration measure $d^2x = \frac{1}{2} du dv$.
At the second equality, we integrate by parts, and finally use $[\mathcal{E}, O] = -2\pi i \d_u O$.
From the condition that $\mathcal{E} \ket{0} = \bra{0} \mathcal{E} = 0$, we reproduce \eqref{dk_corr} and \eqref{dk_corr_R}.

\subsection{A closer look at the partial contact terms}
In this section, we will present an analysis of the partial contact terms in a similar fashion to \cite{HM4d}.
The above discussion used the retarded correlation function because it is in a form that is more convenient for showing the $k$-theorem, but it is equally valid to write the sum rule using the time-ordered correlation function.
Below, we will assume that all correlation functions are time-ordered, and the time order symbol will also be suppressed.
Our interest is to understand the structure of the partial contact terms in the three-point function of the form
\begin{equation}
    \expval{T_{\alpha \beta}(x_3) \d_\mu J_\nu (x_1) \d_\sigma J_\rho(x_2)}
\end{equation}
where $x_3$ approaches $x_1$.
We shall do the OPE expansion around $x_1$, which is different from what we did in the previous subsection (i.e., around $x_3$), but closer to the discussions found in  \cite{HM4d}.

Let us focus on the following form of the partial contact terms
\begin{equation}
    \expval{O^{\sigma_1\dots \sigma_l}(x_1) \d_\sigma J_\rho (x_2)} D_{\alpha \beta \mu \nu \sigma_1 \dots \sigma_l} (\d_3) \delta^{(2)}(x_{31}),
\end{equation}
and in momentum space, such a term will be expressed as (suppressing the overall Dirac delta function for conservation of momentum)
\begin{equation}
    \lll O^{\sigma_1\dots \sigma_l}(q_1 + q_3) \d_\sigma J_\rho(q_2) \rr P_{\alpha\beta\mu\nu \sigma_1\dots \sigma_l}(q_3)
\end{equation}
with $P_{\alpha\beta\mu\nu \sigma_1\dots \sigma_l}(q_3)$ being a polynomial in $q_{3\mu}$ and the metric $g_{\mu \nu}$.
We ignore the $JJ$ partial contact terms because they will not contribute to the sum rule since the ANE operator annihilates the ground state.

Setting $\alpha = \beta = \nu = u$ and $\mu = v$, the partial contact term will become
\begin{equation}
    \lll O^{\sigma_1\dots \sigma_l}(q_1 + q_3) \d_\sigma J_\rho(q_2) \rr P_{uuvu\sigma_1\dots \sigma_l}(q_3).
\end{equation}
In the sum rule, we will eventually put $q_3=0$, so anything with $q_3$ will not contribute to our discussions. It means that $P_{uuvu\sigma_1 \dots \sigma_l}$ must be constructed by only using the metric tensor $g_{uv}$ and the epsilon tensor $\epsilon_{uv}$. For such $P_{uuvu\sigma_1 \dots \sigma_l}$ to be non-zero, the number of $v$ in $\sigma_i$'s is larger than $u$ by two. In other words,  $P_{uuvu\sigma_1 \dots \sigma_l}$ is a projection operator onto the spin 2 component of $O^{\sigma_1\dots \sigma_l}$. Thus, we conclude that any operators that will contribute to the sum rule via the partial contact terms should be spin 2. In our case, we can identify this spin 2 component $O_{uu}  = O^{\sigma_1\dots \sigma_l}P_{uuvu\sigma_1 \dots \sigma_l}$ with $\partial_u J_u$.

What we have learned here is that we cannot exclude the possibility of the partial contact terms in this argument. Indeed, we have seen that the existence of the partial contact term was crucial to derive the correct sum rule.
Note that the appearance of the partial contact terms is slightly different from the previous section due to the fact that we expanded the OPE around a different point, but the contribution is the same.\footnote{More explicitly, we need to reexpress $J(x_3) \partial_3 \delta^{(2)}(x_3-x_1) \sim (J(x_1) + (x_3-x_1) \partial J(x_1) )\partial_3 \delta^{(2)}(x_3-x_1) \sim  J(x_1) \partial \delta^{(2)}(x_3-x_1) - \partial J(x_1) \delta^{(2)}(x_3-x_1) $ as a distribution. The first expression was used in the previous subsection while the last one is used in this subsection.}

Finally, we comment that a similar analysis for the $c$-theorem proof in \cite{HM2d} shows that a spin $2$ and dimension $2$ operator might contribute to the sum rule.
While from just dimensional counting, partial contact terms are ruled out in four dimensions due to unitarity bounds \cite{HM4d}, the same cannot be said for the two-dimensional case -- especially under the presence of the conserved current.\footnote{Indeed, when the conserved current exists, the energy-momentum tensor can mix with $JJ$ or $\partial J$.}
While this does not invalidate the proof when there are no such operators, we may have to reconsider the discussions in the presence of such operators. 

\subsection{The free boson}
Within this and the next subsection, we will verify the sum rule in two theories: the massive boson in this subsection and the massive fermion in the following one.

We will reproduce the expressions for the action, propagator, current, and energy-momentum tensor for convenience. They are the same as in section \ref{boson2}.
The free massive boson action in Euclidean signature is given by
\begin{equation*}
    S_E = \frac{1}{8\pi} \int d^2x \ \d_\mu X \d^\mu X + m^2 X^2 = \frac{1}{4\pi} \int dzd\bar{z} \ (\d X \bar{\d} X + \frac{1}{4}m^2 X^2),
\end{equation*}
The formulas for current and the energy-momentum tensor read
\begin{align}
    J = i \d X ,\ \ 
    T = - \frac{1}{2} \d X \d X.
\end{align}
We have the following expression for the propagator
\begin{align}
    \expval{X(x) X(0)} = 4\pi \int \frac{d^2p}{(2\pi)^2} \frac{e^{ip\cdot x}}{p^2+m^2}
\end{align}
In this case, we can evaluate the separated three-point function as follows
\begin{align}
    &\expval{T(x_3) \bar{\d} J(x_1) \bar{\d} J(x_2)} = \bigg(\frac{m^2}{4}\bigg)^2\expval{\frac{1}{2}\d X \d X(x_3) X(x_1) X(x_2)} 
\end{align}
Here, we used the equation of motion $\bar{\partial} J  = i\bar{\partial} \partial X = i\frac{m^2}{4} X$. 
In the momentum-space computation here, the use of the equation of motion is crucial because it will remove the contact terms that we should not include in the sum rule.
Doing the Fourier transform, one finds
\begin{align}
    \lll T(Q_3)\bar{\d} J(Q_1) \bar{\d} J(Q_2) \rr 
    &= -\pi^2 m^4 \frac{Q_{1z}Q_{2z}}{(Q_1^2+m^2)(Q_2^2+m^2)}. 
\end{align}
In the low energy limit, the above expression becomes $\lll\bar{\d} J(Q_1)\bar{\d} J(Q_2)T(Q_3) \rr \approx - \pi^2 Q_{1z} Q_{2z}$.
This agrees with the sum rule \eqref{dk_mom},
\begin{equation}
    \begin{split}
        k_{\mathrm{UV}}-k_{\mathrm{IR}} 
        &= -\frac{1}{2\pi^2} \bigg(-(\d_{Q_{1z}} -\d_{Q_{2z}})^2 \lll T(Q_3)\bar{\partial} J(Q_1) \bar{\partial} J(Q_2) \rr\bigg) = 1.
    \end{split}
\end{equation}

The use of the equation of motion has the effect of removing the (partial) contact terms.
If we do not do that and perform the Wick contraction straight away, then the result in momentum space will contain contributions from the (partial) contact terms, and read 
\begin{equation*}
    \lll T(Q_3) J(Q_2) J(Q_1) \rr_{\mathrm{QFT}} = -\pi^2
    \frac{Q_{1z}Q_{2z} Q_1^2Q_2^2}{(Q_1^2+m^2)(Q_2^2+m^2)},
\end{equation*}
which vanishes in the sum rule by taking the low momentum limit.
This is consistent with the fact that $k_{\mathrm{IR}} = 0$.
In fact, we can rewrite the momentum-dependent part of the above expression as
\begin{equation*}
    Q_{1z}Q_{2z} \bigg(1 - \frac{m^2}{Q_1^2+m^2} - \frac{m^2}{Q_{2}^2+m^2} + \frac{m^4}{(Q_1^2+m^2)(Q_2^2+m^2)}\bigg)
\end{equation*}
and see how the $\mathrm{UV}$, partial contact, and separated contributions cancel out each other in the sum rule.

\subsection{The free fermion}
Once again, the action for our fermions reads
\begin{equation*}
    S = \frac{1}{2\pi}\int dz d\bar{z} \bigg(\psi^\dagger \bar{\partial} \psi + \tilde{\psi}^\dagger \partial \tilde{\psi} - \frac{m}{2}(\tilde{\psi}^\dagger \psi + \psi^\dagger \tilde{\psi}) \bigg)
\end{equation*}
and the corresponding propagators are given by
\begin{equation*}
        \begin{pmatrix}
        \expval{\psi(x) \tilde{\psi}^\dagger(y)} & \expval{\psi(x) \psi^\dagger(y)} \\
        \expval{\tilde{\psi}(x) \tilde{\psi}^\dagger(y)} & \expval{\tilde{\psi}(x) \psi^\dagger(y)}
    \end{pmatrix}
    = - \int \frac{d^2 p}{2\pi} \frac{e^{ip(x-y)}}{p^2+m^2} 
    \begin{pmatrix}
        m & 2ip_z \\
        2ip_{\bar{z}} & m
    \end{pmatrix}
\end{equation*}
The current and the energy-momentum tensor are given by 
\begin{equation}
        \begin{split}
        J = - \psi^\dagger \psi ,\ \ 
        T = -\frac{1}{2} (\psi^\dagger \partial \psi + \psi \partial \psi^\dagger)
    \end{split}
\end{equation}
Using the equations of motion, we find that
\begin{equation}
    \bar{\partial} J =  - \frac{m}{2} (\psi^\dagger \tilde{\psi} - \tilde{\psi}^\dagger \psi),
\end{equation}
and plugging this into the expression for the separated correlation function, we obtain
\begin{align} \label{3pt_fermion}
    &\expval{\bar{\partial} J(x_1) \bar{\partial} J(x_2) T(x_3)}_{\mathrm{sep}} \nonumber\\
    &=-\frac{m^2}{8} \expval{(\psi^\dagger \tilde{\psi}- \tilde{\psi}^\dagger \psi)(x_1)(\psi^\dagger \tilde{\psi}- \tilde{\psi}^\dagger \psi)(x_2)(\psi^\dagger \partial \psi + \psi \partial \psi^\dagger)(x_3)}
\end{align}
As in the free boson example, if we are working in momentum space, using the equations of motion here is crucial in order to remove the (partial) contact terms that should not be included in the sum rule.
Taking the Fourier transform, we get
\begin{equation} \label{fermion_integral}
\begin{split}
    &\expval{\bar{\d} J(Q_1)\bar{\d} J(Q_2)T(Q_3)} \\
=& \frac{m^2}{2}(2\pi)^3 \int d^2 p \frac{m^2(p_{2z}+p_{3z})(p_{2z}+p_{3z}-p_{1z}) + 4p_{1\bar{z}} p_{2z}p_{3z} (p_{2z}+p_{3z})}{(p_1^2+m^2)(p_2^2+m^2)(p_3^2+m^2)}
\end{split}
\end{equation}
where $p_1 = p, \ p_2 = p + Q_2, \ p_3 = p - Q_1$.
Expanding out the integral in the low-energy limit, one finds 
\begin{equation}
    \lll\bar{\d} J(Q_1)\bar{\d} J(Q_2)T(Q_3) \rr \approx - \pi^2 Q_{1z} Q_{2z},
\end{equation}
which is identical to the bosonic field case; hence, the sum rule holds for this example:  
\begin{equation}
    \Delta k = -\frac{1}{2\pi^2} \bigg(-(\d_{Q_{1z}} -\d_{Q_{2z}})^2 \lll T(Q_3) \bar{\partial} J(Q_1) \bar{\partial} J(Q_2) \rr\bigg) = 1.
\end{equation}
More details on the calculation can be found in the appendix.

Once again, we emphasize that we have applied the equations of motion before doing the Wick contraction.
Without using the equations of motion, the three-point function in momentum space will end up being proportional to $Q_{1\bar{z}} Q_{2\bar{z}}$ and hence drop out in the sum rule.
This is consistent with the fact that $k_{\mathrm{IR}} = 0$.

\section{Summary and Discussion}\label{summary}
Let us recapitulate what we have found. The Lorentzian three-point sum rule for the current central charge  $\Delta k \equiv k_{\mathrm{UV}} - k_{\mathrm{IR}} $ is 
\begin{align*}
    \Delta k &{=}  +\frac{1}{2\pi^2} \int d^2 x_1 d^2 x_2 \delta(u_2) u_1^2 \expval{\mathcal{R} [\mathcal{E}(v=0); \d_v J_u(x_1) \d_v J_u(x_2)]}_{\mathrm{sep}} \\
    & =  -\frac{1}{2\pi^2} \int d^2 x_1 d^2 x_2 \delta(u_2) u_1^2 \expval{\mathcal{T} [\mathcal{E}(v=0) \d_v J_u(x_1) \d_v J_u(x_2)]}_{\mathrm{sep}} 
\end{align*}
in the position space,
or
\begin{align*}
    \Delta k  &  = +\frac{1}{2\pi^2} \bigg((\d_{q_{1u}} -\d_{q_{2u}})^2 \lll\mathcal{R} \ [T_{uu}(q_3); \d_v J_u(q_1) \d_v J_u(q_2) ]\rr_{\mathrm{sep}} \bigg)|_{q_1=q_2=0} \\
    & =-\frac{1}{2\pi^2} \bigg((\d_{q_{1u}} -\d_{q_{2u}})^2 \lll\mathcal{T} \ [T_{uu}(q_3) \d_v J_u(q_1) \d_v J_u(q_2)] \rr_{\mathrm{sep}} \bigg)|_{q_1=q_2=0}    
\end{align*}
in the momentum space. It is free from the (partial) contact term ambiguities. We emphasize that the overall sign is opposite to the naive sum rule without taking into account the partial contact terms. It is remarkable that we have tamed partial contact terms, which are potentially ambiguous, to obtain the unambiguous sum rule.

If we do not worry about the origin of the sign flip, we can just use these correct formulae and apply the positivity of the ANE operator to prove that $\Delta k >0$ under the assumption of the ANEC. We may then declare that the number of charged degrees of freedom decreases under the RG flow due to the positivity of the ANE operator.

There are many ways in which we can generalize our setup.
One interesting generalization is the case with multiple $U(1)$ currents. Throughout the paper, we have restricted ourselves to the case when we have one vectorial (or axial) conserved $U(1)$ current. With multiple currents, even if they are independent in the UV, they might mix in the IR, giving a matrix $k^{\mathrm{IR}}_{ij}$ with elements being $z^4 \expval{J_i(z)J_j(0)}$.

If we repeat our analysis in section 2, we can show that $\dot{K}_{ij} = H_{ij}$ with a positive matrix $H_{ij}$. Obviously, the diagonal elements are monotonically decreasing along the RG flow. 
One interesting question we could ask is what bounds there are on the eigenvalues of $K_{ij}$.
From just the properties of positive definite hermitian matrices alone, we can say that the smallest eigenvalue, along with the sums of ascending eigenvalues, must always decrease.\footnote{This fact is known to the authors as the Schur-Horn theorem.}
It is unclear, however, whether more can be said about each eigenvalue. It is possible some of them can increase, and it would be interesting to see such examples.\footnote{At the same time, it is not obvious if the eigenvalues of $K_{ij}$ are physically important because the diagonalization of $K_{ij}$ may not be physically motivating unless it is compatible with the charge lattice structures.}

In this discussion above, we excluded the case of a nonholomorphic current in the UV.
An example would be the $U(1)$ symmetry of the complex scalar theory.
Written in terms of two real scalars, the action becomes
\begin{equation}
 S =   \frac{1}{8\pi} \int d^2x \ (\partial X_1)^2 + (\partial X_2)^2
\end{equation}
The ``holomorphic" $U(1)$ current, up to a normalization factor, is given by
\begin{equation}
 J_{12} =  X_1 \d X_2 - X_2 \d X_1
\end{equation}
is nonholomorphic.
We can see this explicitly from the fact that the two-point function will have the form
\begin{equation*}
  \langle {J_{12}(z) J_{12}(0)} \rangle =   \frac{k_0 \log |z|^2 + \text{const}}{z^2}.
\end{equation*}
It is unclear whether a $k$-theorem can be formulated for such symmetries. See \cite{Nakamura:2025uyd} for related issues in the $c$-theorem.

In our paper, we have assumed the existence of the $U(1)$ symmetry in the UV, but the symmetry can be emergent or accidental. What can we say about these emergent symmetries from the $k$-thorem? It is interesting to see how the current central charge $k$ can appear. If we try to define ``$k$" by just taking it to be $z^4 \expval{J(z) J(0)}$ where $J$ is the operator that is conserved in the IR but not conserved in the UV, we see that this quantity diverges in the UV limit $z \to 0$.
This is because in the UV, $J$ is not conserved and it has the scaling dimension $(h, \bar{h}) = (1+\gamma_J,  \gamma_J)$, where $\gamma_J$ is positive in unitary theories.
Therefore, our $K$ function behaves like $\frac{1}{(z\bar{z})^{2\gamma_J}}$ in the UV limit. This means that for emergent symmetries, the current central charge is infinite in the UV and approaches the conformal value $k$ in the IR limit. For such emergent symmetries, we have infinitely charged degrees of freedom in the UV!

Finally, the most ambitious generalizations will be the new sum rule and the analogue of the $k$-theorem in higher dimensions. We recall that the ANEC-based proof put the $c$ and $a$ theorem on an equal footing \cite{HM2d,HM4d,Hartman:2024xkw}.
We hope our new proof of the $k$-theorem will inspire the search for the analogous theorem in higher dimensions. Some higher $d$ reference based on two-point functions can be found in \cite{Vilasis-Cardona:1994oke,Karateev:2020axc,Baume:2024poj}. In supersymmetric field theories, the $\tau_{RR}$ minimization principle may be close to the idea of ``$k$-theorem" \cite{Barnes:2005bm, Buican:2011ty}. It is often the case that the supersymmetry relates the energy-momentum tensor central charges to the R-current central charges, and it may be a good starting point to discuss the monotonicity.

\section*{Acknowledgements}
YN is in part supported by JSPS KAKENHI Grant Number 21K03581. UN is grateful for the scholarship from the Kyoto iUP Program.

\appendix

\section{Detail of the fermion calculation}
Here, we shall give more details on how the fermion correlation function is calculated.
Expanding out \eqref{3pt_fermion}, we get
\begin{equation}
    -\frac{m^2}{8} (\text{[I]} + \text{[II]} - \text{[III]} )
\end{equation}
where
\begin{equation}
    \text{[I]} = -\expval{\psi^\dagger \tilde{\psi}(x_1) \psi^\dagger \tilde{\psi}(x_2) \partial \psi \psi^\dagger (x_3)} + \expval{\psi^\dagger \tilde{\psi}(x_1) \psi^\dagger \tilde{\psi}(x_2)  \psi \partial \psi^\dagger (x_3)}
\end{equation}
\begin{equation}
    \text{[II]} = -\expval{\tilde{\psi}^\dagger \psi(x_1) \tilde{\psi}^\dagger \psi(x_2) \partial \psi \psi^\dagger (x_3)} + \expval{\tilde{\psi}^\dagger \psi(x_1) \tilde{\psi}^\dagger \psi(x_2)  \psi \partial \psi^\dagger (x_3)}
\end{equation}
\begin{equation}
    \begin{split}
        \text{[III]} &= -\expval{\tilde{\psi}^\dagger \psi (x_1) \psi^\dagger \tilde{\psi}(x_2) \partial \psi \psi^\dagger(x_3)} - \expval{\psi^\dagger \tilde{\psi}(x_1) \tilde{\psi}^\dagger \psi (x_2) \partial \psi \psi^\dagger(x_3)} \\
        &+\expval{\tilde{\psi}^\dagger \psi (x_1) \psi^\dagger \tilde{\psi}(x_2)  \psi \partial\psi^\dagger(x_3)} + \expval{\psi^\dagger \tilde{\psi}(x_1) \tilde{\psi}^\dagger \psi (x_2) \partial \psi \partial \psi^\dagger(x_3)}.
    \end{split}
\end{equation}
Doing Wick contraction and plugging in the formulas for the propagators, we get
\begin{align}
    \text{[I]} = \text{[II]} &= - \int \frac{d^2p_1 d^2p_2 d^2p_3}{(2\pi)^3} \frac{e^{ip_1x_{12} + ip_2 x_{23} + ip_3 x_{31}}}{(p_1^2+m^2)(p_2^2+m^2)(p_3^2+m^2)} 2m^2(p_{2z}+p_{3z})^2 \\
    \text{[III]} &=  2\int \frac{d^2 p_1 d^2p_2 d^2p_3}{(2\pi)^3} \frac{e^{ip_1 x_{12} + i p_2 x_{23} + ip_3 x_{31}}}{(p_1^2+m^2)(p_2^2+m^2)(p_3^2+m^2)} \\ &\times\bigg(2m^2 p_{1z} (p_{2z} + p_{3z}) - 8p_{3z} p_{2z} p_{1\bar{z}} (p_{2z}+p_{3z}) \bigg)
\end{align}
Adding everything together and doing a Fourier transform, we can get the expression \eqref{fermion_integral}.
The integral itself can be done as follows.
Using Feynman's trick
\begin{equation*}
    \begin{split}
    &\frac{1}{(p_1^2+m^2)(p_2^2+m^2)(p_3^2+m^2)} \\
    =&2 \int dx dy dz \ \delta(x+y+z-1) \frac{1}{(x(p_1^2+m^2)+y(p_2^2+m^2)+z(p_3^2+m^2))^3},
    \end{split}
\end{equation*}
the integral becomes
\begin{equation}
    \begin{split}
    &\frac{m^2}{2}(2\pi)^3 \int d^2 p \times 2 \int dx dy dz  \ \delta(x+y+z-1) \\
    \times & \frac{1}{(p^2 + m^2 +xQ_1^2 +yQ_2^2 - (yQ_2-xQ_1)^2)^3} \\
    \times& \big( m^2(2p_z + 2xQ_{1z} -2Q_{2z} - Q_{1z} + Q_{2z}) (p_z + x Q_{1z} - yQ_{2z}-Q_{1z}+Q_{2z}) \\
    &+ 4 (p_{\bar{z}} + xQ_{1\bar{z}}-yQ_{2\bar{z}}) (p_z + x Q_{1z}-yQ_{2z}+Q_{2z}) \\ & \times (p_z + xQ_{1z}-yQ_{2z}-Q_{1z})(2p_{z}+2xQ_{1z}-2yQ_{2z}-Q_{1z}+Q_{2z}) \big)
    \end{split}
\end{equation}
After doing so, the $Q_1,Q_2$ in the denominator shoud not contribute to the final sum rule so one can approximate it as 
\begin{equation}
    \frac{1}{(p^2+m^2)^3}.
\end{equation}
Doing the $x, y$ integral we get
\begin{equation}
\begin{split}
    &\int_0^1 dx \int_0^{1-x} dy (2p_z + 2xQ_{1z} -2Q_{2z} - Q_{1z} + Q_{2z}) \\ & \times(p_z + x Q_{1z} - yQ_{2z}-Q_{1z}+Q_{2z}) \\ &= \dots + \frac{1}{6}(Q_{1z}^2 - Q_{1z}Q_{2z}+Q_{2z}^2) \\
    &\int_0^1 dx \int_0^{1-x} dy (p_z + x Q_{1z}-yQ_{2z}+Q_{2z}) \\ &\times (p_z + xQ_{1z}-yQ_{2z}-Q_{1z})(2p_{z}+2xQ_{1z}-2yQ_{2z}-Q_{1z}+Q_{2z}) \\ &= \dots -\frac{1}{2}Q_{1z}Q_{2z} p_{z}
\end{split}
\end{equation}
The integrals above contain many other terms, but here we will only focus on terms that are quadratic in $Q_{1z}, Q_{2z}$ since only such terms will contribute to the sum rule in the end.
Using rotational symmetry (that is replacing $p_{\bar{z}} p_z$ with $\frac{1}{4}p^2$)
\begin{equation}
    \begin{split}
    &\expval{\bar{\d} J(Q_1)\bar{\d} J(Q_2) T(Q_3)}_{\mathrm{sep}} \\  =&\frac{m^2}{2}(2\pi)^3 \delta^{(2)}(Q_1+Q_2+Q_3) \int d^2 p \frac{\frac{m^2}{3}(Q_{1z}^2-Q_{1z}Q_{2z}+Q_{2z}^2)- p^2Q_{1z}Q_{2z}}{(p^2+m^2)^3}
    \end{split}
\end{equation}
In the sum rule, since one can set $Q_{3z}$ to zero,  it is possible to make the replacement $Q_{1z}^2-Q_{1z}Q_{2z}+Q_{2z}^2 \to -3Q_{1z}Q_{2z}$.
Doing the integral gives 
\begin{equation}
    (2\pi)^2 \delta^{(2)}(Q_1+Q_2+Q_3) \bigg(-\pi^2 Q_{1z} Q_{2z}\bigg),
\end{equation}
which is what we claimed above.
%%%%%%%%%%%%%%%%%%%%%%%%%%%%%%%%%%%%%%%%%%%%%%%%%%%%%%%%
%%%%%%%%%%%%%%%%%%%%%%%%%%%%%%%%%%%%%%%%%%%%%%%%%%%%%%%%
% bibliography via BibTeX
\bibliographystyle{JHEP}
\bibliography{main}

%%%%%%%%%%%%%%%%%%%%%%%%%%%%%%%%%%%%%%%%%%%%%%%%%%%%%%%%
%%%%%%%%%%%%%%%%%%%%%%%%%%%%%%%%%%%%%%%%%%%%%%%%%%%%%%%%

\end{document}